\documentclass[aps,prd,preprint,superscriptaddress,tightenlines,twocolumn,nofootinbib,showpacs,floatfix]{revtex4}

\usepackage{graphicx} 
\usepackage{dcolumn}  
\usepackage{bm}       

\begin{document}

\newcommand{\BR}{{\cal B}(\Upsilon({\rm 1S})\to\gamma f_2'(1525))=(4.0\pm1.3\pm0.6)\times 10^{-5}}

\preprint{CLNS 10/2072}  

\preprint{CLEO 10-09}    

\title{\Large \boldmath $\Upsilon({\rm 1S})\to\gamma f_2'(1525);~f_2'(1525)\to K^0_{\rm S}K^0_{\rm S}$ decays}

\author{D.~Besson}
\author{D.~P.~Hogan}
\altaffiliation[Now at: ]{University of California Berkeley, Berkeley, CA 94720}
\affiliation{University of Kansas, Lawrence, Kansas 66045, USA}
\author{T.~K.~Pedlar}
\affiliation{Luther College, Decorah, Iowa 52101, USA}
\author{D.~Cronin-Hennessy}
\author{J.~Hietala}
\author{P.~Zweber}
\affiliation{University of Minnesota, Minneapolis, Minnesota 55455, USA}
\author{S.~Dobbs}
\author{Z.~Metreveli}
\author{K.~K.~Seth}
\author{A.~Tomaradze}
\author{T.~Xiao}
\affiliation{Northwestern University, Evanston, Illinois 60208, USA}
\author{S.~Brisbane}
\author{L.~Martin}
\author{A.~Powell}
\author{P.~Spradlin}
\author{G.~Wilkinson}
\affiliation{University of Oxford, Oxford OX1 3RH, UK}
\author{H.~Mendez}
\affiliation{University of Puerto Rico, Mayaguez, Puerto Rico 00681}
\author{J.~Y.~Ge}
\author{D.~H.~Miller}
\author{I.~P.~J.~Shipsey}
\author{B.~Xin}
\affiliation{Purdue University, West Lafayette, Indiana 47907, USA}
\author{G.~S.~Adams}
\author{D.~Hu}
\author{B.~Moziak}
\author{J.~Napolitano}
\affiliation{Rensselaer Polytechnic Institute, Troy, New York 12180, USA}
\author{K.~M.~Ecklund}
\affiliation{Rice University, Houston, Texas 77005, USA}
\author{J.~Insler}
\author{H.~Muramatsu}
\author{C.~S.~Park}
\author{L.~J.~Pearson}
\author{E.~H.~Thorndike}
\author{F.~Yang}
\affiliation{University of Rochester, Rochester, New York 14627, USA}
\author{S.~Ricciardi}
\affiliation{STFC Rutherford Appleton Laboratory, Chilton, Didcot, Oxfordshire, OX11 0QX, UK}
\author{C.~Thomas}
\affiliation{University of Oxford, Oxford OX1 3RH, UK}
\affiliation{STFC Rutherford Appleton Laboratory, Chilton, Didcot, Oxfordshire, OX11 0QX, UK}
\author{M.~Artuso}
\author{S.~Blusk}
\author{R.~Mountain}
\author{T.~Skwarnicki}
\author{S.~Stone}
\author{J.~C.~Wang}
\author{L.~M.~Zhang}
\affiliation{Syracuse University, Syracuse, New York 13244, USA}
\author{G.~Bonvicini}
\author{D.~Cinabro}
\author{A.~Lincoln}
\author{M.~J.~Smith}
\author{P.~Zhou}
\author{J.~Zhu}
\affiliation{Wayne State University, Detroit, Michigan 48202, USA}
\author{P.~Naik}
\author{J.~Rademacker}
\affiliation{University of Bristol, Bristol BS8 1TL, UK}
\author{D.~M.~Asner}
\altaffiliation[Now at: ]{Pacific Northwest National Laboratory, Richland, WA 99352}
\author{K.~W.~Edwards}
\author{K.~Randrianarivony}
\author{G.~Tatishvili}
\altaffiliation[Now at: ]{Pacific Northwest National Laboratory, Richland, WA 99352}
\affiliation{Carleton University, Ottawa, Ontario, Canada K1S 5B6}
\author{R.~A.~Briere}
\author{H.~Vogel}
\affiliation{Carnegie Mellon University, Pittsburgh, Pennsylvania 15213, USA}
\author{P.~U.~E.~Onyisi}
\author{J.~L.~Rosner}
\affiliation{University of Chicago, Chicago, Illinois 60637, USA}
\author{J.~P.~Alexander}
\author{D.~G.~Cassel}
\author{S.~Das}
\author{R.~Ehrlich}
\author{L.~Fields}
\author{L.~Gibbons}
\author{S.~W.~Gray}
\author{D.~L.~Hartill}
\author{B.~K.~Heltsley}
\author{D.~L.~Kreinick}
\author{V.~E.~Kuznetsov}
\author{J.~R.~Patterson}
\author{D.~Peterson}
\author{D.~Riley}
\author{A.~Ryd}
\author{A.~J.~Sadoff}
\author{X.~Shi}
\author{W.~M.~Sun}
\affiliation{Cornell University, Ithaca, New York 14853, USA}
\author{J.~Yelton}
\affiliation{University of Florida, Gainesville, Florida 32611, USA}
\author{P.~Rubin}
\affiliation{George Mason University, Fairfax, Virginia 22030, USA}
\author{N.~Lowrey}
\author{S.~Mehrabyan}
\author{M.~Selen}
\author{J.~Wiss}
\affiliation{University of Illinois, Urbana-Champaign, Illinois 61801, USA}
\author{J.~Libby}
\affiliation{Indian Institute of Technology Madras, Chennai, Tamil Nadu 600036, India}
\author{M.~Kornicer}
\author{R.~E.~Mitchell}
\author{C.~M.~Tarbert}
\affiliation{Indiana University, Bloomington, Indiana 47405, USA }
\collaboration{CLEO Collaboration}
\noaffiliation

\date{\today}

\begin{abstract}
We report on a study of exclusive radiative decays of the $\Upsilon$(1S) resonance into a
final state consisting of a photon and 
two $K^0_{\rm S}$ candidates.
We find evidence for 
a signal for $\Upsilon$(1S)$\to\gamma f_2'$(1525); $f_2'$(1525)$\to\gamma K^0_{\rm S}K^0_{\rm S}$,
at a rate $\BR$,
consistent with previous observations
of $\Upsilon$(1S)$\to\gamma f_2'$(1525); $f_2'$(1525)$\to K^+K^-$, and isospin. 
Combining this branching fraction with
existing branching fraction measurements of
$\Upsilon$(1S)$\to\gamma f_2'(1525)$ and
$J/\psi\to\gamma f_2'(1525)$, we obtain the ratio of
branching fractions:
${\cal B}(\Upsilon{\rm(1S)}\to\gamma f_2'(1525))/{\cal B}(J/\psi\to\gamma f_2'(1525))=0.09\pm0.02$, approximately consistent with
expectations based on soft collinear effective theory.
\end{abstract}
\smallskip

\maketitle

A particularly interesting class of $\Upsilon$(1S) decays are the
radiative decays, which could show evidence for the same type
of two-body resonance production
as has been observed in $\psi$ decay. 
The most naive arguments simply scale the
charge-dependence of the coupling and the mass dependence of
the propagator in the associated amplitude, leading
to bottomonium/charmonium radiative widths varying as
[($q_{b}/q_{c}$)($m_{c}/m_{b}$)]$^{2}\approx 1/36$. The
ratio of
the full widths of the (1S) charmonium vs.\ bottomonium states
(93 keV/54 keV)~\cite{PDG10} implies radiative
bottomonium branching fractions approximately 4--5\% of that of
the corresponding charmonium state. This naive expectation is
consistent with measurements of radiative
decays into spin-zero mesons (e.g., $\gamma\eta(')$),
although considerably smaller than measurements for
decays into spin-two mesons (e.g., $\gamma f_2$).

A comprehensive calculation
using soft-collinear effective theory (SCET)
and non-relativistic QCD has been
implemented to calculate the ratio of `non-exotic' branching fractions
${\cal B}(\Upsilon{\rm{(1S)}}\to\gamma f_2)/{\cal B}(J/\psi\to\gamma f_2)$~\cite{SCET}.
That theory calculation gives
a predicted ratio of (0.13--0.18), slightly larger than the currently measured value for the
$f_2'$(1525) ($0.08\pm0.03$~\cite{PDG10}), but 
not inconsistent with extant data, given the large errors.
The CLEO Collaboration has
previously presented results on
exclusive radiative decays into two charged tracks~\cite{luis}, as well as the
final states $\gamma\pi^0\pi^0$ and $\gamma\eta\eta$~\cite{holger}.
We now supplement those 
measurements and searches with a study of decays into
a photon plus two $K^0_{\rm S}$, with $K^0_{\rm S}\to\pi^+\pi^-$.

The CLEO~III detector was 
operated as a general purpose solenoidal magnet spectrometer and
calorimeter. Approximately 10 ${\rm fb}^{-1}$ of data were collected in the region of
the $\Upsilon$(4S), supplemented by ~1 ${\rm fb}^{-1}$ samples of data around each of the narrow,
lower-mass resonances. 
The analysis described herein is
based on a sample of 21.2 million $\Upsilon$(1S) events, plus
10.2 million events taken on the continuum, just below the
$\Upsilon$(4S) resonance.

Elements of the detector, as well as performance characteristics relevant to this analysis
are described in detail elsewhere ~\cite{r:CLEO-II,r:CLEOIIIa,r:CLEOIIIb}. 
Particularly important in defining the candidate signal sample for this
signal topology is photon detection and energy resolution.
For photons in the central
``barrel'' region of the CsI 
electromagnetic calorimeter, at energies
greater than 2 GeV, the energy resolution is approximately 1--2\%.
The tracking system
used to identify the charged pion candidates, the 
RICH particle identification system, 
and the electromagnetic
calorimeter are all contained within a 1~Tesla superconducting
coil.
Neutral $K^0_{\rm S}$ candidates are identified by CLEO's
standard reconstruction software as oppositely-signed
charged pion pairs with a common origin point away from the primary 
vertex and have an invariant mass within 12 ${\rm MeV/c^2}$ of the
nominal $K^0_{\rm S}$ mass. Dipion candidates within 
24 ${\rm MeV/c^2}$ of the nominal $K^0_{\rm S}$ mass
are defined as ``sideband'' $K^0_{\rm S}$ candidates and
are retained for background evaluation.

To obtain our
candidate event sample, we select those events
containing four
charged tracks (with total charge zero) 
that combine to form two $K^0_{\rm S}$ candidates. We
allow a maximum of one `extra' charged track 
in the event, which is ignored in subsequent analysis.
Each $K^0_{\rm S}$ candidate must have an invariant mass within three
units of the experimental mass resolution of
the nominal $K^0_{\rm S}$ mass, corresponding to approximately 12 ${\rm MeV/c^2}$.
Charged pion $K^0_{\rm S}$ decay candidates
are required to have $dE/dx$ information consistent with
that expected for charged pions, within 3 standard deviations in energy 
deposition resolution.
To suppress possible QED contamination,
we require that
the four charged tracks must be inconsistent with
an $e^+e^-\to\tau\tau$ ``1-prong vs.\ 3-prong''
charged-track topology and also have no charged track 
positively identified as an electron or muon.
Beyond the inner tracking chambers, we require 
one high-energy electromagnetic shower observed in the
barrel calorimeter which does not match 
(within 0.1 radians) the position of any charged
track extrapolated beyond the drift chamber into the barrel calorimeter.
Finally, the sum of the observed photon energy plus the energies
of the drift chamber tracks (assumed to be pions)
must lie within 120 MeV
(roughly, 2.5 standard deviations) of
the total center of mass energy. The magnitude of
the total event momentum 
must be within 120 MeV/c of the expected value of zero, as well.

\begin{figure}[htpb]
\centerline{\includegraphics[width=0.5\textwidth]{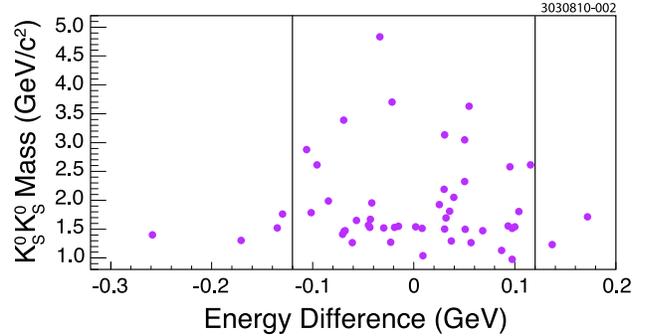}}
\caption{$K^0_{\rm S}K^0_{\rm S}$ invariant mass vs.\ (Total visible energy -- center-of-mass energy) for events
satisfying overall momentum conservation. Acceptance region is bounded
by vertical lines.}
\label{fig:dE_dp}
\end{figure}

For our event candidates,
we observe a cluster of events that conserve overall four-momentum with
an approximate energy difference resolution of 100 MeV, 
as shown in
the invariant mass vs.\ energy difference plot (Fig.\ \ref{fig:dE_dp}).

\begin{figure}[htpb]
\centerline{\includegraphics[width=0.5\textwidth]{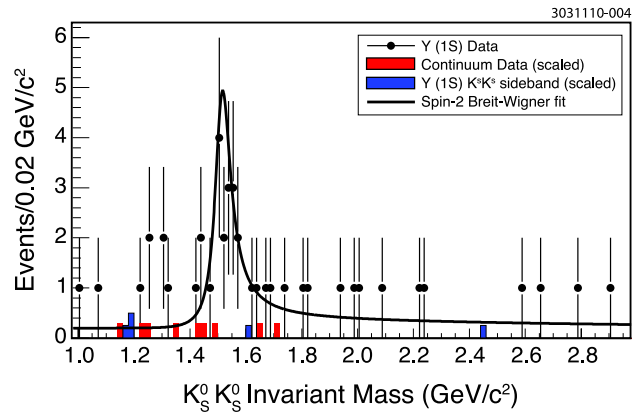}}
\caption{Invariant mass of $K^0_{\rm S}K^0_{\rm S}$ candidates for events satisfying all
energy, momentum, and photon selection requirements, showing signal as well as background estimators from the continuum and also $K^0_{\rm S}$ sidebands. 
Also overlaid is the fit to the relativistic, 
spin-2 Breit-Wigner signal shape.}
\label{fig:f2fit}
\end{figure}

After imposing energy and momentum conservation requirements,
the $f_2'(1525)\to K^0_{\rm S}K^0_{\rm S}$ candidate 
signal
is shown in Fig.\ \ref{fig:f2fit}. 
We note the absence of any signal in events selected from either
$K^0_{\rm S}K^0_{\rm S}$ sidebands, or data taken from the 
continuum in the vicinity of the $\Upsilon$(4S) resonance.
Extrapolated to the resonant $\Upsilon$(1S) sample,
we can attribute a maximum of 
two of the observed resonant events to the under-lying
continuum, with no obvious peaking under the $f_2'(1525)$.
Defining the $K^0_{\rm S}$ sidebands as the region from 
$0.12\to0.24$ ${\rm GeV/c^2}$ from the
nominal $K^0_{\rm S}$ mass, we obtain an extrapolated yield of $\approx$2
such sideband contributions 
in the entire $K^0_{\rm S}K^0_{\rm S}$ invariant mass interval.
We scale this value by 
a factor of 1/8 to extrapolate the sideband yield to the
signal, giving a maximum net contribution of $<$0.4 potential signal events.

To ensure that the observed signal is not a mis-reconstruction of
the known decay $\Upsilon({\rm 1S})\to\gamma 4\pi$, we have run our reconstruction
code on a sample of simulated
Monte Carlo $\Upsilon({\rm 1S})\to\gamma 4\pi$ events, statistically
equivalent to the number expected in data, for which
the 4 pions are distributed 
according to a simplistic phase space model. Doing so, 
we observe 3 events which are reconstructed as $K^0_{\rm S}K^0_{\rm S}$, with
no peaking in the candidate signal region.
In general, asymmetric $\pi^0$ decays can lead to a topology with
a highly energetic photon and a much smaller energy photon which can go
undetected. This leads to concerns about possible contamination from
hadronic decays of the type
$\Upsilon({\rm 1S})\to\pi^0 f_2'(1525)$. However, this decay violates
C-parity and therefore cannot contribute to the background.

We have fit the candidate signal, after applying all candidate and event
selection requirements to a relativistic, spin-2 Breit-Wigner signal plus
a flat background (Fig. \ref{fig:f2fit}). 
The likelihood 
fit yield, with mass and width
constrained to the PDG values
(M=$(1525\pm5)$ MeV and
$\Gamma=(73\pm6)$ MeV, respectively~\cite{PDG10})
corresponds to $N_{sig}=16.6\pm5.3$ signal events. 
Inclusion of possible
$f_2(1270)\to K^0_{\rm S}K^0_{\rm S}$ and 
$f_0(1710)\to K^0_{\rm S}K^0_{\rm S}$ 
components gives yields for those two resonances
statistically consistent with zero and results in a variation in the central
value for the $f_2'(1525)$ signal of less than 4\%.
The efficiency for the decay chain $\Upsilon({\rm 1S})\to\gamma f_2'(1525)$;
$f_2'(1525)\to K^0_{\rm S}K^0_{\rm S}$ is assessed with 10,000 dedicated 
Monte Carlo simulated events, and estimated to be $18.5\pm0.4$\% (statistical
error only), not including
branching fractions.

Systematic errors are estimated as follows: a) photon-finding
efficiency uncertainty (2\%),
b) $K^0_{\rm S}K^0_{\rm S}$ detection efficiency (8\%), 
c) total number of $\Upsilon$(1S) events (2\%), 
d) efficiency uncertainty due to component branching
fraction errors and limited Monte Carlo statistics (4\%),
and e) fitting systematics.
This last systematic uncertainty is determined as follows:
the difference between the area found using a relativistic, spin-2 Breit-Wigner in data 
(our default parametrization) is 7\%
smaller in data with parameters fixed according to the Particle Data Group $f_2'(1525)$ parameters vs.\
floated parameters. The difference between using a second-order vs.\ a first-order Chebyschev polynomial
background results in an additional 9\% variation in fitted area. As mentioned above,
adding possible $\Upsilon({\rm 1S})\to\gamma f_2(1270)$ 
and 
$\Upsilon({\rm 1S})\to\gamma f_0(1710)$ structure to our fit changes the fitted 
$f_2'(1525)$ area by less than 4\%.
Taken together in quadrature,
we assess a total systematic uncertainty of 14\% (relative).

We translate our fit yield into a branching fraction by knowing
${\cal B}(f_2'\to K{\overline K})=(0.888\pm0.031)$, the fraction of 
$K{\overline K}$ which is $K^0_{\rm S}K^0_{\rm S}$ (1/4), the
branching fraction
${\cal B}(K^0_{\rm S}\to\pi^+\pi^-)=(0.6920\pm0.0005)$, 
and the
Monte Carlo efficiency of 18.5\%, giving a total efficiency of
$\epsilon_{tot}=
(0.888\pm0.031)\times
0.25\times
(0.6920\pm0.0005)^2\times
(0.185\pm0.004)$. Combining our signal yield of $N_{sig}$ events and
the
total efficiency (($19.7\pm0.7)\times 10^{-4}$) with the
total number of $\Upsilon$(1S) events (21.2$\times 10^6$) yields a final branching fraction
estimate of $\BR$,
compared with the 
previous
CLEO branching fraction 
measurement of $(3.7^{+0.9}_{-0.7}\pm0.8)\times 10^{-5}$, based on the
$\gamma K^+K^-$ final state~\cite{luis}. 
Comparing the likelihood of the fit result to the
likelihood obtained when the signal yield is set to zero,
we find $-2\ln(\Delta\cal{L})=4.0$, a significance of $4.0\sigma$. 
In this expression, $\Delta \cal{L}$ is the difference in likelihood between the two fits.
Within errors, we find
good agreement between the values derived from the 
charged vs.\ neutral kaon decay modes.

In summary, we have observed exclusive radiative decays of the $\Upsilon$(1S) meson
into the $\gamma K^0_{\rm S}K^0_{\rm S}$ final state. A large 
$f_2'$(1525) signal is observed in the di-$K^0_{\rm S}$ mass spectrum, 
with a
branching fraction $\BR$, consistent with previous measurements
of $\Upsilon({\rm 1S})\to\gamma f_2'$(1525); $f_2'(1525)\to K^+K^-$.
Although no predictions for this final state, per se, exist in the literature,
we can nevertheless compare our calculated branching fraction,
relative to the analogous branching fraction for $J/\psi$ decays, with
the predictions from SCET~\cite{SCET}. Combining our current result with the
previous result for $\Upsilon({\rm 1S})\to\gamma f_2'(1525)\to\gamma K^+K^-$, we obtain
an updated estimate ${\cal B}(\Upsilon\to\gamma f_2'(1525))=(3.8\pm0.9)\times 10^{-5}$.
The ratio of experimental branching
fractions: 
$R_2\equiv{\cal B}(\Upsilon({\rm 1S})\to\gamma f_2)/{\cal B}(J/\psi\to\gamma f_2)=0.09\pm0.02$ for the
$f_2'(1525)$, 
consistent with both the experimental results for the $f_2(1270)$ ($R_2=0.071\pm0.008)$, as well as
the predictions of SCET. The equality of these ratios for the $f_2(1270)$ and
the $f_2'(1525)$ is consistent with the naive expectation from SU(3) symmetry.


\begin{acknowledgments}
We gratefully acknowledge the effort of the CESR staff
in providing us with excellent luminosity and running conditions.
D.~Cronin-Hennessy thanks the A.P.~Sloan Foundation.
This work was supported by
the National Science Foundation,
the U.S. Department of Energy,
the Natural Sciences and Engineering Research Council of Canada, and
the U.K. Science and Technology Facilities Council.
\end{acknowledgments}


\begin{thebibliography}{99}
\bibitem{PDG10} K. Nakamura {\it et al.} (Particle Data Group), J. Phys. G {\bf 37}, 075021 (2010).
\bibitem{SCET}S. Fleming, C. Lee, and A. Leibovich, Phys. Rev. D {\bf 71}, 074002 (2005). 
\bibitem{luis}S.B. Athar {\it et al.} (CLEO Collaboration), Phys. Rev. D {\bf 73}, 032001 (2006).
\bibitem{holger}D. Besson {\it et al.} (CLEO Collaboration), Phys. Rev. D {\bf 75}, 072001 (2007).
\bibitem{r:CLEO-II}D.~G. Cassel {\it et al.}, Nucl. Inst. Meth. A {\bf 252}, 325 (1986).
\bibitem{r:CLEOIIIa}D. Peterson {\it et al.}, Nucl. Inst. Meth. A {\bf 478}, 142 (2002).
\bibitem{r:CLEOIIIb}M. Artuso {\it et al.}, Nucl. Inst. Meth. A {\bf 502}, 91 (2003).
\end{thebibliography}
\end{document}